\documentclass[twocolumn,aps,prc,superscriptaddress,showpacs]{revtex4}
\usepackage{amsmath,bm}%
\usepackage{graphicx}%

\begin{document}
\title{Cluster emission and phase transition behaviours in nuclear disassembly }
\author{Y. G. Ma }
%\thanks{}
%\email{ygma@cyclotronmail.tamu.edu}

\affiliation{Shanghai Institute of Nuclear Research, Chinese
Academy of Sciences, P.O. Box 800-204, Shanghai 201800, CHINA}
%\footnotemark \footnotetext{}
\affiliation{China Center of Advanced Science and Technology
(World Laboratory), P. O. Box 8730, Beijing 100080, CHINA}
%\affiliation{LPC, IN2P3-CNRS, ISMRA et Universit\'e, Boulevard Mar\'echal Juin,
%14050 Caen Cedex, FRANCE }
\affiliation{Cyclotron Institute, Texas A\&M
 University, College Station, Texas 77843-3366, USA}

%\date{\today}
\date{J. Phys. G: Nucl. Part. Phys. {\bf{27}} (2001) 2455-2470.}
\begin{abstract}

The features of the emissions of light particles (LP), charged particles (CP), 
intermediate mass fragments (IMF) and the largest fragment (MAX) are 
investigated for $^{129}Xe$ as functions of temperature and 'freeze-out' density 
in the frameworks of the isospin-dependent lattice gas model and the classical 
molecular dynamics model. Definite turning points for the slopes of average 
multiplicity of LP, CP and IMF, and of the mean mass of the largest fragment 
( $A_{max}$ ) are shown around a liquid-gas phase transition temperature and while 
the largest variances of the distributions of LP, CP, IMF and MAX appear 
there. It indicates that the cluster emission rate can be taken as a probe of 
nuclear liquid--gas phase transition. Furthermore, the largest fluctuation is 
simultaneously accompanied at the point of the phase transition as can be 
noted by investigating both the variances of their cluster multiplicity or mass 
distributions and the Campi scatter plots within the lattice gas model and the 
molecular dynamics model, which is consistent with the result of the traditional 
thermodynamical theory when a phase transition occurs. 
\end{abstract}
\pacs{ 25.70.Pq, 05.70.Jk, 24.10.Pa, 02.70.Ns}

\maketitle

\section{Introduction}

Phase transition and critical phenomenon is an extensively debatable subject in the natural 
sciences. Recently, the same concept was introduced into the astronomical objects [1] and 
the microscopic systems, such as in atomic cluster [2] and nuclei [3], of these the nuclei, 
as a microscopic finite-size system, are attracting more and more nuclear experimentalists 
to search for the liquid--gas phase transition and investigate their behaviour. To date, 
various experimental evidences have cumulated which seem to be related to the nuclear 
phase transition. For instance, violent heavy-ion collisions break the nuclei into several 
intermediate mass fragments, which can be viewed as a critical phenomenon as observed 
in fluid, atomic and other systems. It prompts a possible signature on the liquid--gas phase 
transition in the nuclear system. The sudden opening of the nuclear multifragmentation 
and vaporization [4] channels can be interpreted as the signature of the boundaries of phase 
mixture [5]. In addition, the plateau of the nuclear caloric curve in a certain excitation energy 
range gives a possible indication of a first-order phase transition [6, 7] as predicted in the 
framework of statistical equilibrium models [8]. On the other hand, the extraction of critical 
exponents in the charge or mass distribution of the multifragmentation system [9] can be 
explained as an evidence of phase transition. More recently, the negative microcanonical 
heat capacity was experimentally observed in nuclear fragmentation [10] which relates to the 
liquid--gas phase transition [11], and in atomic cluster [12] which relates to solid to liquid phase 
transition [13], respectively. Moreover, some evidence of spinodal decomposition in nuclear 
multifragmentation was recently obtained experimentally [14], which shows the presence of 
liquid--gas phase coexistence region and gives a strong argument in favour of the existence of 
first-order liquid--gas phase transition in finite nuclear systems. 

Meanwhile, several theoretical models have been developed to treat such a phase 
transition in the nuclear disassembly, e.g. percolation model, lattice gas model, statistical 
multifragmentation model and molecular dynamics model etc (e.g. see the recent review 
article of Richert and Wagner [15] and references therein). In this paper, we are interested in 
the lattice gas model ( LGM ) and classical molecular dynamics ( CMD ) model [16]. The former 
is a simple short range interaction model [17], but it can successfully be applied to nuclear 
systems with isospin symmetry and asymmetry. LGM is carried assuming that the system is 
in a 'freeze-out' density $\rho_f$ with thermal equilibrium at temperature T. Previous calculations 
[18] with LGM showed that there exists a phase transition for the finite nuclear systems by 
studying the effective power law parameter ( $\tau$ ) of cluster mass or charge distribution, their 
second moments ( $S_2$ ) and the specific heat. More recently, we proposed two novel criteria, 
namely multiplicity information entropy ( $H$ ) and nuclear Zipf's law to diagnose the onset 
of liquid--gas phase transition in the framework of the isospin-dependent LGM (I-LGM) and 
isospin-dependent classical molecular dynamics (I-CMD) model [19]. 

In this paper, we show that the emission rate of clusters is a useful tool to diagnose the 
nuclear liquid--gas phase transition, while the largest fluctuation of cluster multiplicities is 
simultaneously revealed at the point of the phase transition by investigating the features of 
light particles (LP), charged particles (CP), intermediate mass fragments (IMF) and the largest 
fragment (MAX) of disassembling source in I-LGM and I-CMD frameworks. 

The paper is organized as follows. Section 2 gives the descriptions of I-LGM and I-CMD. 
Results and discussions are shown in section 3 where the multiplicities of cluster emissions, 
their slopes and their fluctuations are investigated. The influence of the 'freeze-out' density 
on cluster emission is also presented in the framework of I-LGM and I-CMD and the role 
of Coulomb interaction is studied by comparing the results of I-LGM and I-CMD at a given 
'freeze-out' density. Finally, the conclusion is given in section 4. 

\section{Description of models} 

\subsection{ Isospin-dependent lattice gas model}
 
The lattice gas model of Lee and Yang [17], where the grand canonical partition function of 
a gas with one type of atoms is mapped into the canonical ensemble of an Ising model for 
spin 1/2 particles, has successfully described the liquid--gas phase transition for the atomic 
system. The same model has already been applied to the microscopic nuclear system, e.g. see 
the papers of Pan and Das Gupta et al [16]. In order to better understand the context of this 
study, the models are described below. 

In the LGM, A ( = N + Z ) nucleons with an occupation number s which is defined as the 
`spin' s = 1 ( -1 ) for a proton (neutron) or s = 0 for a vacancy, are placed in the L sites of a, 
three-dimensional cubic lattice. Each cubic lattice has a size 1.0/$\rho_0$ = 6.25 $fm^3$ and can, at 
most, be occupied by a single nucleon, where $\rho_0$ = 0.16 $fm^{-3}$ is the normal nucleon density. 
Nucleons in the nearest neighbouring sites interact with an 
energy s i s j. The Hamiltonian of the system is written as 
\begin{equation}
 H = \sum_{i=1}^{A} \frac{P_i^2}{2m} - \sum_{i < j} \epsilon_{s_i s_j}s_i s_j .
\end{equation}
The interaction constant  $\epsilon_{s_i s_j}$ is related to the binding energy of the nuclei. In order to 
incorporate the isospin effect in the LGM, the short-range interaction 
constant $\epsilon_{s_i s_j}$  
is chosen 
to be different between the nearest neighbouring like nucleons and unlike nucleons, 
\begin{eqnarray}
 \epsilon_{nn} \ &=&\ \epsilon_{pp} \ = \ 0. MeV \nonumber , \\
 \epsilon_{pn} \ &=&\ - 5.33 MeV,
\end{eqnarray}
which indicates the repulsion between the nearest-neighbouring like nucleons and attraction 
between the nearest-neighbouring unlike nucleons. This kind of isospin-dependent interaction 
incorporates, to a certain extent, the Pauli exclusion principle and effectively avoids producing 
unreasonable clusters, such as di-proton and di-neutron clusters etc. The disassembly of the 
system is calculated at an assumed `freeze-out' density $\rho_f$ = (A/L)$\rho_0$ , beyond $\rho_f$ nucleons 
are too far apart to interact. 

In this model, N + Z nucleons are put in L sites by Monte Carlo sampling using the 
canonical Metropolis algorithm [20]. As pointed out in [21], however, one has to be careful 
treating the process of Metropolis sampling in order to satisfy the detailed balance principle 
and therefore warrant the correct equilibrium distribution in the final state. Describing in detail, 
in this paper, first an initial configuration with N + Z nucleons is established. Second, for 
each event, a sufficient number of 'spin'-exchange steps are tested, e.g. 20 000 steps to let the 
system generate states with a probability proportional to the Boltzmann probability distribution 
with the Metropolis algorithm. In each 'spin'-exchange step, a random trial change on the 
basis of the previous configuration is made. For instance, we choose a nucleon at random 
and attempt to exchange it with one of its neighbouring nucleons or vacancies regardless of 
the sign of its 'spin' (Kawasaki-like spin-exchange dynamics [22]), then compute the change 
$\Delta E$ in the energy of the system due to the trial change. If $\Delta$E is less than or equal to zero, 
accept the new configuration and repeat the next 'spin'-exchange step. If $\Delta E$ is positive, 
compute the 'transition probability' W = $e^{-\Delta E/T}$ and compare it with a random number r 
in the interval [0, 1]. If r $\leq$  W , accept the new configuration, otherwise retain the previous 
configuration. Twenty thousand 'spin'-exchange steps are performed to ensure that we get that 
the equilibrium state (afterwards we will show the 'spin'-xchange step dependence of some 
observables). Third, once the nucleons have been placed stably on the cubic lattice after 20000 
'spin'-exchange steps for each event, their momenta are generated by Monte Carlo sampling 
of the Maxwell--Boltzmann distribution. Thus various observables based on phase space can 
be calculated in a straightforward fashion for each event. One important point of such Monte 
Carlo Metropolis computations is that the above 'spin'-exchange approach between the nearest 
neighbors, independently of their 'spin', is evidenced to be satisfied by the detailed balance 
condition as noted in [21]. In other words, this sampling method guarantees that the generated 
microscopic states form an equilibrium canonical ensemble. 
 
Once this is done the LGM immediately gives the cluster distribution by using the rule 
that two nucleons are part of the same cluster if their relative kinetic energy is insufficient to 
overcome the attractive bond [18]: 
\begin{equation}
P_r^2/2\mu - \epsilon_{s_i s_j}s_i s_j < 0.
\end{equation}
This method has been proved to be similar to the so-called Coniglio--Klein prescription in 
condensed matter physics [23] and was shown to be valid in LGM. 

\subsection{Isospin-dependent classical molecular dynamics model }

Since the LGM is a model of the nearest-neighbouring interaction, a long-range Coulomb 
force is not amenable to lattice gas type calculation. Pan and Das Gupta [16, 17] provide 
a prescription, based on simple physical reasoning, to decide if two nucleons, occupying 
neighbouring sites form part of the same cluster or not [26]. They first try to map the 
LGM calculation to a molecular dynamics type prediction, both first carried out without any 
Coulomb interaction. If the calculations match quite faithfully, then they can study the effects 
of the Coulomb interaction by adding that to the molecular dynamics calculation. Here 
we adopt the same prescription to use the molecular dynamics and therefore investigate the 
Coulomb effect. The results and conclusions are now compared and checked between the LGM 
and the CMD.Obviously, here we do not perform any ab initio molecular dynamics calculation 
but only use it for simulating the nuclear disassembly, starting from a thermally equilibrated 
source which has been produced by the above I-LGM: i.e. the nucleons are initialized at 
their lattice sites with Metropolis sampling and have their initial momenta with Maxwell-- 
Boltzmann sampling. From this starting point we switch the calculation to CMD evolution 
under the influence of a chosen force. Note that in this case $\rho_f$ is, strictly speaking, not a 
`freeze-out' density for molecular dynamics calculation but merely defines the starting point 
for time evolution. However, since classical evolution of a many particle system is entirely 
deterministic and the initialization does have in it all the information of the asymptotic cluster 
distribution, we continue to call $\rho_f$ the 'freeze-out' density. 

The form of the force in the CMD is also chosen to be isospin dependent in order to 
compare with the results of I-LGM. The potential for unlike nucleons is expressed as [16, 24] 
\begin{widetext}
\begin{eqnarray}
 v_{\rm n p}(r) (r/r_0<a)\ &=&\ C\left[B(r_0/r)^p-(r_0/r)^q\right]\nonumber
    exp({\frac{1}{(r/r_0)-a}}), \\
v_{\rm  n p}(r) (r/r_0>a)\ &=&\ 0.
\label{pot}
\end{eqnarray}
\end{widetext}
where $r_0$ = 1.842 fm is the distance between the centres of two adjacent cubes so that 
$\rho_0$ = 1/${r_0}^3$ = 0.16 $fm^{-3}$. The parameters of the potentials are p = 2, q = 1, a = 1.3, B = 
0.924 and C = 1966 MeV. With these parameters the potential is minimum at $r_0$ with the value 
-5.33 MeV, zero when the nucleons are more than 1.3$r_0$ apart and strongly repulsive when 
r is significantly less than $r_0$ . We now turn to the nuclear potential between like nucleons. 
Although we take $\epsilon_{pp}$ = $\epsilon_{nn}$ = 0 in I-LGM, 
the fact that we do not put two like nucleons in 
the same cube suggests that there is short-range repulsion between them. We have taken the 
nuclear force between two like nucleons to be the same expressions as above +5.33 MeV up 
to r = 1.842 fm and zero afterwards, 
\begin{eqnarray}
v_{\rm p p}(r) (  r < r0 )\ &=&\  v_{\rm n p}(r)- v
_{\rm n p}(r_0)\nonumber , \\
v_{\rm p p}(r) ( r > r0 )\ &=&\ 0.
\end{eqnarray}
Figure 1 shows the above potential $v_{np}$ or $v_{pp}$. This potential form automatically cuts off at 
r/$r_0$ = a (equation (4)) or r/$r_0$ = 1 ( equation (5) ) without discontinuities in any r derivatives, 
which is a distinct advantage in any molecular dynamics simulation application. 

\begin{figure}
\includegraphics[scale=0.43]{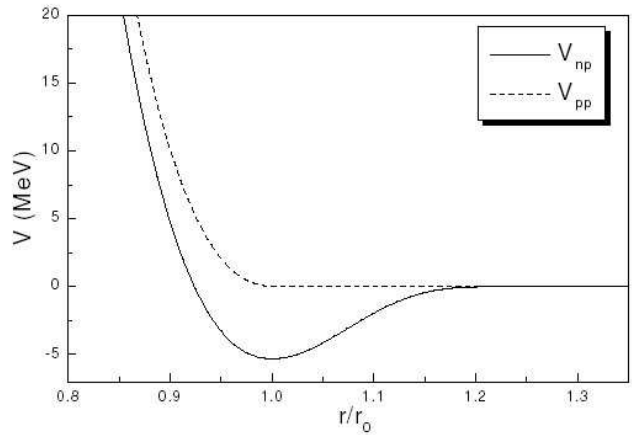}
\caption{\footnotesize Molecular dynamics potential for like nucleon pair ( $v_{pp}$ ) 
and unlike nucleon pair ( $v_{np}$ ).  }
\label{fig1}
\end{figure}

The system evolves with the above potential. The time evolution equations for each 
nucleon are, as usual, given by 
\begin{eqnarray}
 \partial \vec{p_i}/\partial
t\ &=&\ - \Sigma_{j\neq i} \bigtriangledown _i v(r_{ij})\nonumber , \\
\partial \vec{r_i}/\partial t\ &=&\ \vec{p_i}/m.
\end{eqnarray}
Numerically, the particles are propagated in the phase space by a well-known Verlet 
algorithm [25], one of the finite-difference methods in molecular dynamics with continuous 
potentials. At asymptotic times, for instance, the original blob of matter expands to 64 times 
its volume in the initialization, the clusters are easily recognized: nucleons which stay together 
after an arbitrarily long time are part of the same cluster. The observables based on cluster 
distribution in both models are now compared while they are also compared by switching 
on/off the Coulomb interaction within the molecular dynamics.

\section{ Results and discussions}
 
We choose a medium-size nucleus $^{129}Xe$ as an example. The input parameters are temperature 
T and 'freeze-out' density $\rho_f$ in the model calculations. In this study T is mostly limited to 
the range of 1.5--7 MeV and the 'freeze-out' density $\rho_f$ changes in a wide range, namely, 
0.097$\rho_0$, 0.18$\rho_0$, 0.38$\rho_0$ and 0.60$\rho_0$, which corresponds to the total 
cubic lattices $11 \times 11 \times 11$, $9 \times 9 \times 9$, $7 \times 7 \times 7$ and
$6 \times 6 \times 6$, respectively, of which, 0.60$\rho_0$ is the maxmium 
freeze-out density which is allowed for $^{129}Xe$ because a cubic lattice is required in our LGM 
calculations. One thousand events are simulated for each combination of T and $\rho_f$ which 
ensures good statistics for results. 

Before we present the results, we would like to check the role of the 'spin'-exchange 
step and from that we can know how the system tends to the equilibrium state in the model. 
Suppose, we wish to determine experimentally the value of a property of a system such as 
internal energy or cluster emission. In general, such properties depend upon the phase space 
of A nucleons. Over time, the instantaneous value of the property fluctuates as a result of 
interactions between the nucleons. However, when the system reaches its equilibrium state, 
the average of the instantaneous values over huge samples (`ensemble average') are viewed 
as experimental asymptotic values. But when will the system approach its equilibrium in 
the framework of LGM? To this end, we use the time average of the instantaneous values to 
investigate this question. In the LGM Monte Carlo simulation, the 'spin'-exchange step is 
viewed as a time step. We display some step-averaged obsevables that evolve with the step. 
Figure 2 shows the step-averaged total kinetic energy per nucleon $E_{kin}$, total potential energy 
per nucleon $E_{pot}$, the multiplicity information entropy $H$ which is defined in section 3.1., the 
multiplicities of emitted neutrons, protons, CP, IMF, and the mass of the heaviest fragment at 
$\rho_f$ = 0.38$\rho_0$ and T= 3 MeV or 6 MeV in I-LGM. From this figure we see that the observables 
tend to their asymptotic values after 1000 steps, i.e. the system tends to equilibrium after 1000 
'spin'-exchange steps in the lattice gas Monte Carlo simulation. In the following results, we 
adopt the results after 20 000 'spin'-exchange steps when the system is in the equilibrium 
state. 

\begin{figure}
\includegraphics[scale=0.43]{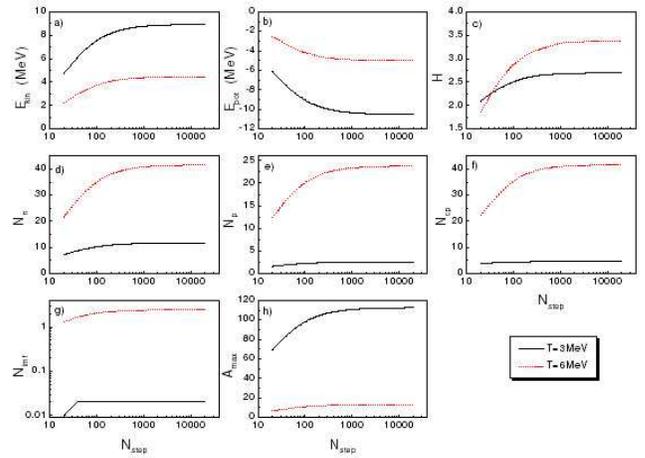}
\caption{\footnotesize The averaged total kinetic energy per nucleon $E_{kin}$ (a), total potential energy per nucleon 
$E_{pot}$ (b), the multiplicity information entropy $H$ (c), the multiplicities of the emitted neutrons $N_n$ 
(d), protons $N_p$ (e), charged particles $N_{cp}$ (f) and intermediate mass fragment $N_{imf}$ (g), the heaviest 
fragment mass $A_{max}$ (h) over the 'spin'-exchange step as a function of step at $\rho_f$ = 0.38 $\rho_0$
 and T = 3 or 6 MeV in the framework of I-LGM. The symbols are illustrated in the right bottom. }
\label{fig2}
\end{figure}

\begin{figure}
\includegraphics[scale=0.43]{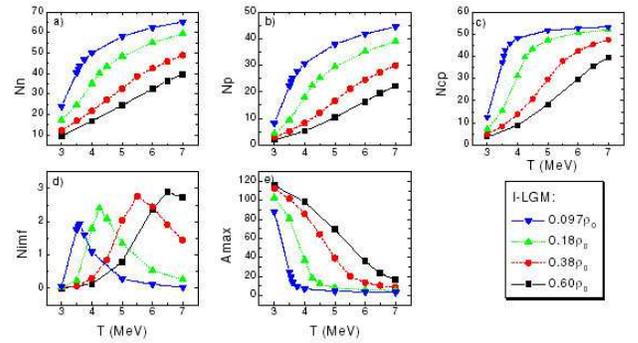}
\caption{\footnotesize  Average multiplicity of the emitted neutrons $N_n$ (a), 
protons $N_p$ (b), charged particles $N_{cp}$ 
(c) and intermediate mass fragment $N_{imf}$ (d), the mean mass of the largest fragment $A_{max}$ (e) as a 
function of temperature in different `freeze-out' density in the framework of I-LGM. The symbols 
are illustrated in the right bottom. }
\label{fig3}
\end{figure}

\subsection{I-LGM in different 'freeze-out' densities}

Figure 3 shows that the mean multiplicities of emitted neutrons, protons, CP and IMF and the 
mean mass for the largest fragment evolve with temperature at different `freeze-out' densities 
in the I-LGM calculation. Here IMF is defined as 3 $\leq$ Z $\leq$ 20. At a fixed `freeze-out' density, 
average neutron multiplicity ( $N_n$ ), proton multiplicity ( $N_p$ ), charged particle multiplicity ($N_{cp}$ ) 
and the mean mass for the largest fragment ( $A_{max}$ ) display monotonous increase or decrease 
with temperature as expected. But the multiplicity ( $N_{imf}$ )ofIMFshowsariseandfallwith 
temperature [27, 28], when the system probably crosses the phase transition boundary. With 
decreasing `freeze-out' density, $N_p$, $N_n$, $N_{cp}$ and $A_{max}$ increase, since larger space separation 
among nucleons at smaller `freeze-out' density makes the clusters less bound and therefore 
the sizes of free clusters decrease and then the cluster multiplicities increase. The situation of 
$N_{imf}$ is slightly complicated, i.e. it increases with the decrease of `freeze-out' density in the 
lower temperature branch contrary to the higher temperature branch. 

It seems difficult to discover the possibility of phase transition of nuclei if we only see 
these mean quantities as shown above ( $N_{imf}$ is an exception ). However, when we focus on their 
slopes to temperature (figure 4), sharp changes are observed at nearly the same temperature 
at each fixed `freeze-out' density, for instance, 3.5 MeV at 0.097$\rho_0$, 4MeV at 0.18$\rho_0$, 5MeV 
at 0.38$\rho_0$ and 6 MeV at 0.60$\rho_0$. At such a transition point, (1) the multiplicities of emitted 
clusters increase rapidly and after that the emission rate slows down; and (2) the decrease in the 
largest fragment size reaches a valley for such a finite system. Physically, the largest fragment 
is simply related to the order parameter $\rho_l - \rho_g$ (the difference of density in nuclear `liquid' 
and `gas' phases). In infinite matter, the infinite cluster exists only on the `liquid' side of the 
critical point. In finite matter, the largest cluster is present on both sides of the phase transition 
point. In this calculation, a valley for the slope of $A_{max}$ to temperature may correspond to a 
sudden disappearance of infinite cluster (`bulk liquid') near the phase transition temperature. It 
is not the occasional production of such waves of the slopes; it should reflect the onset of phase 
transition there. This idea is supported by surveying the other phase transition observables, 
such as the effective power law parameter $\rho_f$rom the mass or charge distribution of fragment 
and the information entropy $H$ of event multiplicity distribution [19]. $H$ can be expressed as 
\begin{equation}
H = - \sum_{i \in Event} p_i \cdot ln(p_i)
\end{equation}
where $p_i$ is defined as the event-normalized total multiplicity probability and 
$\sum_{i \in Event} p_i$ = 1. 
Figure 5 depicts these results. The minima of $\tau$ and the maxima of $H$ appear around respective 
phase transition temperatures at different values of `freeze-out' density, i.e. about 3.5 MeV 
at 0.097$\rho_0$ , 4--4.25 MeV at 0.18$\rho_0$, 5--5.5 MeV at 0.38$\rho_0$ and 6--6.5 MeV at 0.60$\rho_0$.
These temperatures are consistent with those extracted from the choppy position of the above slopes. 
It indicates that the above slopes (emission rate) can be taken as a probe of a liquid--gas phase 
transition of nuclei. 

\begin{figure}
\includegraphics[scale=0.43]{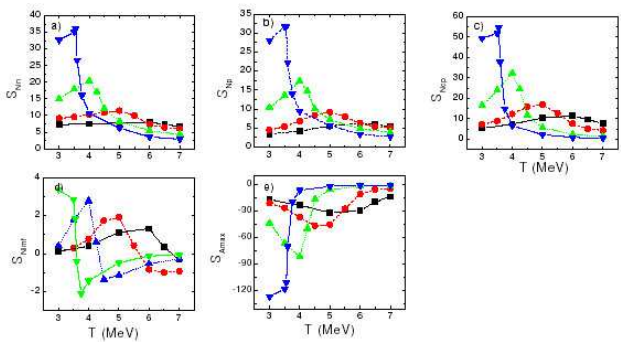}
\caption{\footnotesize Same as in figure 3, but for their slopes with temperture. The symbols
are the same as in figure 3.}
\label{fig4}
\end{figure}

Furthermore, the largest fluctuations of cluster multiplicities are found around the phase 
transition point in the same calculation. Figure 6 illustrates that RMS width ( $\sigma$ ) of the 
multiplicity distributions of neutrons, protons, CP and IMF, and the distribution of the 
largest fragment masses. These variances generally show peaks at the same phase transition 
temperatures as those extracted from the above observables for each fixed `freeze-out' density. 
Note that the fluctuation of $A_{max}$ is related to the compressibility of the system. These features 
are also consistent with one of the phase transition behaviours, i.e. the largest fluctuation at 
the phase transition point exists [29]. This fluctuation represents an internal feature of the 
disassembling system, not a numerical fluctuation. 

Another way to characterize the fluctuation is the method of the conditional moments 
introduced by Campi [30]. The normalized second moment $S_2$ in each event is defined as 
\begin{equation}
S_2 = \frac{ \sum_{ A_i \neq A_{max}} {A_i}^2 \cdot n_i (A_i) }{\sum_{ A_i \neq A_{max}} {A_i} \cdot n_i (A_i)}
\end{equation}
where $n_i$ is the multiplicity of cluster mass $A_i$ , and the summation is over all clusters in an 
event except the heaviest one which corresponds to the bulk liquid in an infinite system. For 
example, we show the 1000-event Campi scatter plots ( ln($A_{max}$) versus ln($S_2$) ) as a function 
of temperature at a fixed density of 0.38$\rho_0$ in figure 7. Actually, the plots clearly illustrate 
the evolution of the disassembling mechanism with temperature. At lower temperatures, 
only the under-critical (liquid phase) branch with a negative slope of ln($A_{max}$) versus ln($S_2$) 
exists while at higher temperature, only the super-critical (gas phase) branch with a positive 
slope of ln($A_{max}$) versus ln($S_2$) appears. However, both the branches (the liquid--gas phase 
coexistence region) meet closely around 5.5 MeV, which indicates the onset of the liquid--gas 
phase transition [28, 30, 31]. Similar behaviour shows for the calculation at 0.097$\rho_0$ ,0.18$\rho_0$ 
and 0.60$\rho_0$ at their respective phase transition temperature. 

We point out that the `freeze-out' density-dependent phase transition temperature, 
extracting from $\tau$ and $S_2$ , was also observed in a previous study [32]. Similar to water, 
the temperature of liquid--gas phase transition decreases with pressure. In the nuclear case, the 
decrease in `freeze-out' density is similar to the decrease in the internal pressure inside nuclei, 
hence it leads to decrease in transition temperature. But this $\rho_f$-dependent phenomenon 
vanishes when excitation energy is used as a variable in the LGM [33]. In other words, the 
excitation energy has perhaps a good correspondence with the critical temperature
because of only one critical point and hence only one critical temperature for a system.

\begin{figure}
\includegraphics[scale=0.43]{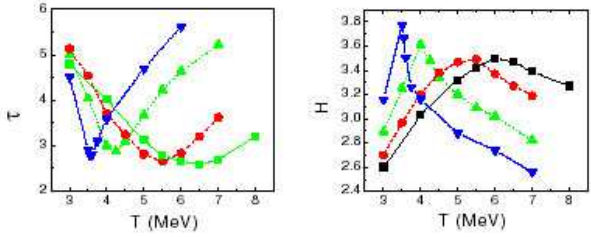}
\caption{\footnotesize The effective power law parameter $\tau$ of cluster mass 
distribution (left) and the information 
entropy $H$ (right) as a function of temperature in different `freeze-out' 
densities in the framework 
of I-LGM. The symbols are the same as in figure 3.}
\label{fig5}
\end{figure}

\begin{figure}
\includegraphics[scale=0.43]{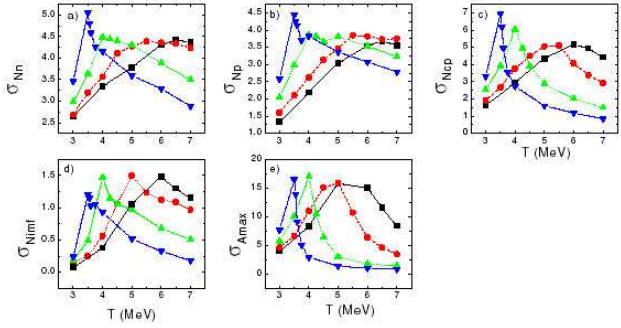}
\caption{\footnotesize Temperature-dependent RMS widths ( $\sigma$ ) of 
the distributions of $N_n$ (a), 
$N_p$ (b), $N_{cp}$ (c), 
$N_{imf}$ (d) and $A_{max}$ (e) in different `freeze-out' densities 
in the framework of I-LGM. The symbols 
are the same as in figure 3. }
\label{fig6}
\end{figure}

\begin{figure}
\includegraphics[scale=0.43]{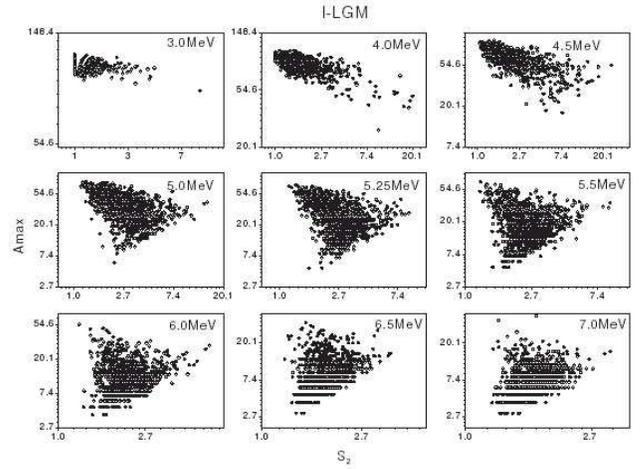}
\caption{\footnotesize Campi scattering plots for I-LGM at 0.38$\rho_0$ . 
excitation energy has perhaps a good correspondence with the critical temperature because of 
only one critical point and hence only one critical temperature for a system.}
\label{fig7}
\end{figure}

\subsection{Roles of Coulomb force: comparison of I-LGM and I-CMD in fixed 'freeze-out' density }

Considering the absence of long-range Coulomb force in the LGM, we adopt the CMD to 
investigate the Coulomb effect and check the features of cluster emission and its relation to 
phase transition behaviour. To this end, first we make a comparison for the results of I-CMD 
with Coulomb or without Coulomb and those of I-LGM at a certain fixed `freeze-out' density, 
namely 0.38$\rho_0$ . Second, we present the results of all these observables in the frame of I-CMD 
with Coulomb at four different `freeze-out' densities to check cluster emission and its relation 
to the phase transition behaviour in next subsection. 

Figure 8 shows that $N_p$ , $N_n$ , $N_{cp}$ , $N_{imf}$ and $A_{max}$ change with temperature in different 
calculation cases, i.e. I-CMD with Coulomb or without Coulomb and I-LGM (see the meaning 
of the symbols in the figure.). The multiplicities of clusters and the largest fragment mass are 
close to each other between I-LGM and I-CMD with Coulomb except for the multiplicities 
of neutrons and protons, illustrating that the I-LGM is, in general, a good tool to describe the 
fragmentation if Coulomb interaction can be ignored. When Coulomb interaction is switched 
on, $N_p$ , $N_{cp}$ and $N_{imf}$ increase due to the repulsive role among protons while $A_{max}$ decreases. 
Meanwhile, $N_n$ does not change because of no Coulomb interaction. $N_{imf}$ also shows a rise 
and fall with temperature in the I-CMD cases. Coulomb force makes the turning temperature 
of $N_{imf}$ smaller due to long-range repulsion.

\begin{figure}
\includegraphics[scale=0.43]{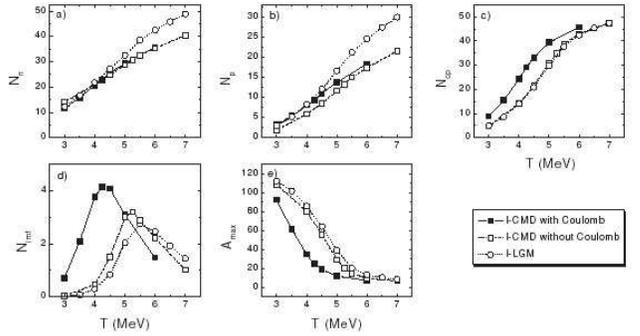}
\caption{\footnotesize Same as in figure 3, but for comparison between different calculations: I-LGM (open 
circles), I-CMD without Coulomb interaction (open squares) and I-CMD with Coulomb force 
(solid squares). The `freeze-out' density of system is 0.38$\rho_0$ . }
\label{fig8}
\end{figure} 

\begin{figure}
\includegraphics[scale=0.43]{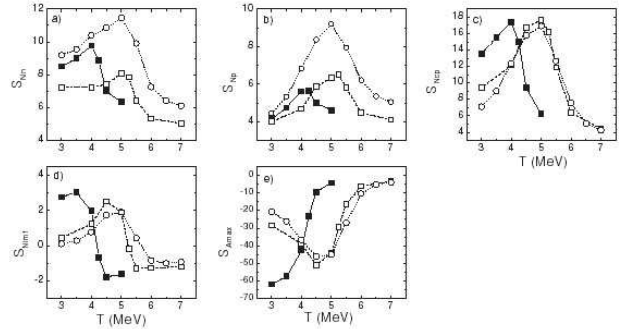}
\caption{\footnotesize Same as in figure 4, but for comparison between different calculations: I-LGM (open 
circles), I-CMD without Coulomb interaction (open squares) and I-CMD with Coulomb force 
(solid squares). The `freeze-out' density of system is 0.38$\rho_0$ . }
\label{fig9}
\end{figure}   

The slopes of multiplicities of emitted clusters and of mean mass of the largest fragment 
are plotted as a function of temperature in case of I-CMD in figure 9. The definite peaks 
of slopes are found as in the I-LGM case. The corresponding temperature at the peaks is 
located about 4 MeV in the I-CMD case with the Coulomb interaction and about 5 MeV in 
the I-CMD case without the Coulomb interaction. This turning temperature also reflects the 
onset of phase transition there. If we investigate $\tau$ and $H$ (figure 10), we find that there are a 
minima of $\tau$ and the maxima of $H$ around phase transition temperatures, i.e. about 4--4.25 MeV 
for I-CMD with the Coulomb, around 5 MeV for I-CMD without the Coulomb and around 
5.5 MeV for I-LGM respectively. 

Finally, the RMS widths of the multiplicity distributions of the clusters and of the largest 
fragment mass are checked in the I-CMD case in figure 11. The widths of the multiplicity 
distributions of neutrons and protons tend to be saturated at higher temperature, while those 
for CP, IMF and $A_{max}$ demonstrate peaks at a certain fixed temperature, i.e. around 4 MeV 
for the case I-CMD with the Coulomb, 5 MeV for the case of I-CMD without the Coulomb, 
which is similar to the I-LGM case. These turning temperatures are also consistent with the 
phase transition temperature as shown in figures 8--10 in the I-CMD cases. 

Overall, Coulomb interaction plays a notable role in favour of fragment production and 
reduces the temperature of the phase transition. When the system is small, the Coulomb 
interaction is not expected to be important and I-LGM could be a good tool to treat nuclear 
disassembly. But for large nuclear systems, the neglect of the Coulomb force in LGM is rather 
serious handicap. In this background, first, we made a suitable selection for the molecular 
dynamics interaction potential and then got good agreement between the I-LGM and I-CMD 
without the Coulomb. Thanks to this agreement, we treated the nuclear disassembly more
realistically with switching on of the Coulomb interaction in the frame of I-CMD afterwards.

\begin{figure}
\includegraphics[scale=0.43]{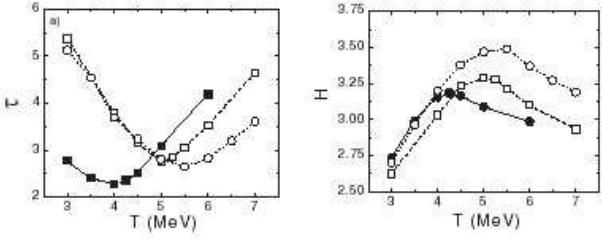}
\caption{\footnotesize Same as in figure 5, but for comparison between different calculations: I-LGM (open 
circles), I-CMD without Coulomb interaction (open squares) and I-CMD with Coulomb force 
(solid squares). The `freeze-out' density of system is 0.38$\rho_0$ . }
\label{fig10}
\end{figure}  

\begin{figure}
\includegraphics[scale=0.43]{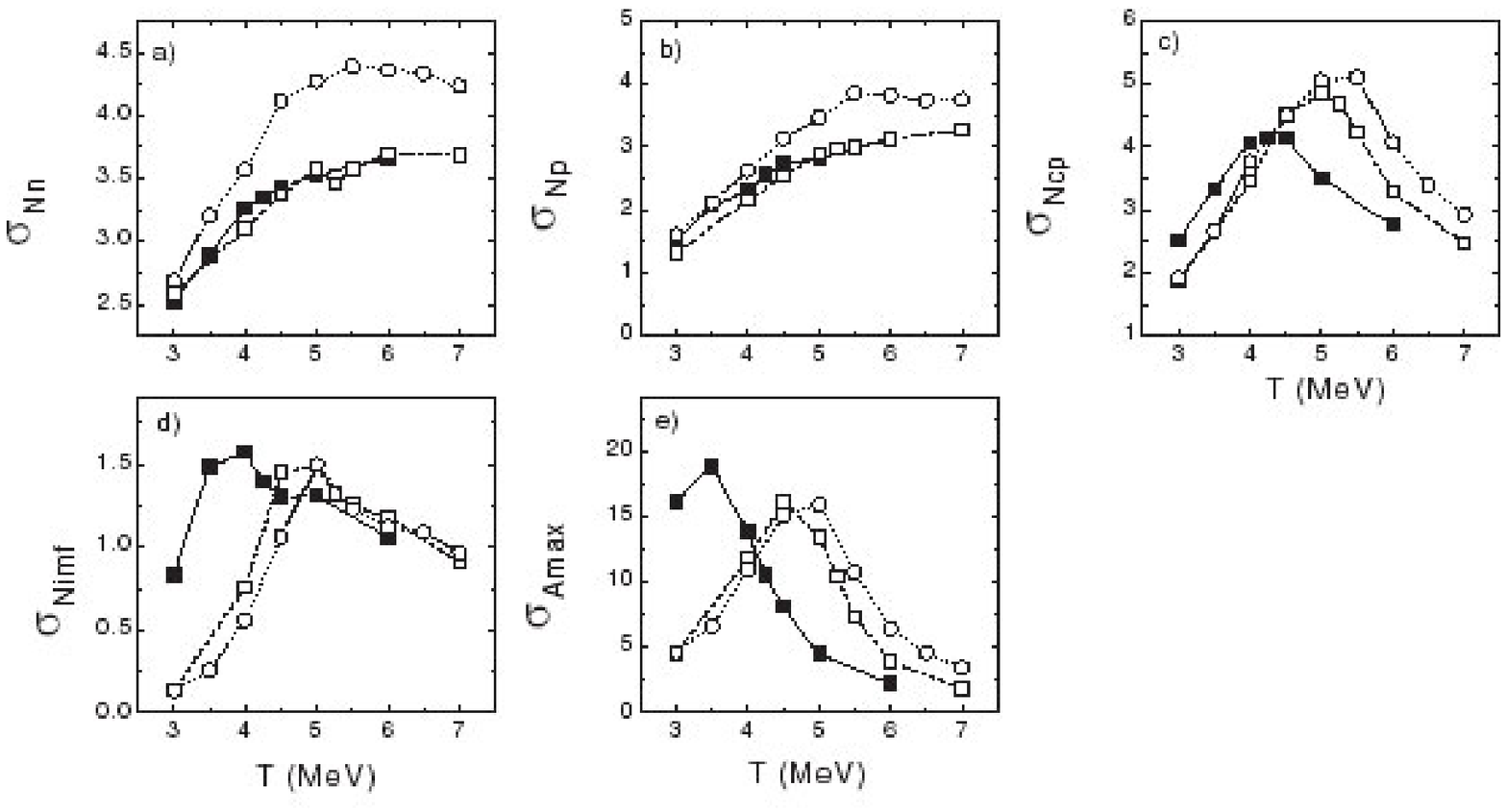}
\caption{\footnotesize Same as in figure 6, but for comparison between different calculations: I-LGM (open 
circles), I-CMD without Coulomb interaction (open squares) and I-CMD with Coulomb force 
(solid squares). The `freeze-out' density of system is 0.38$\rho_0$ . }
\label{fig11}
\end{figure}  

\begin{figure}
\includegraphics[scale=0.43]{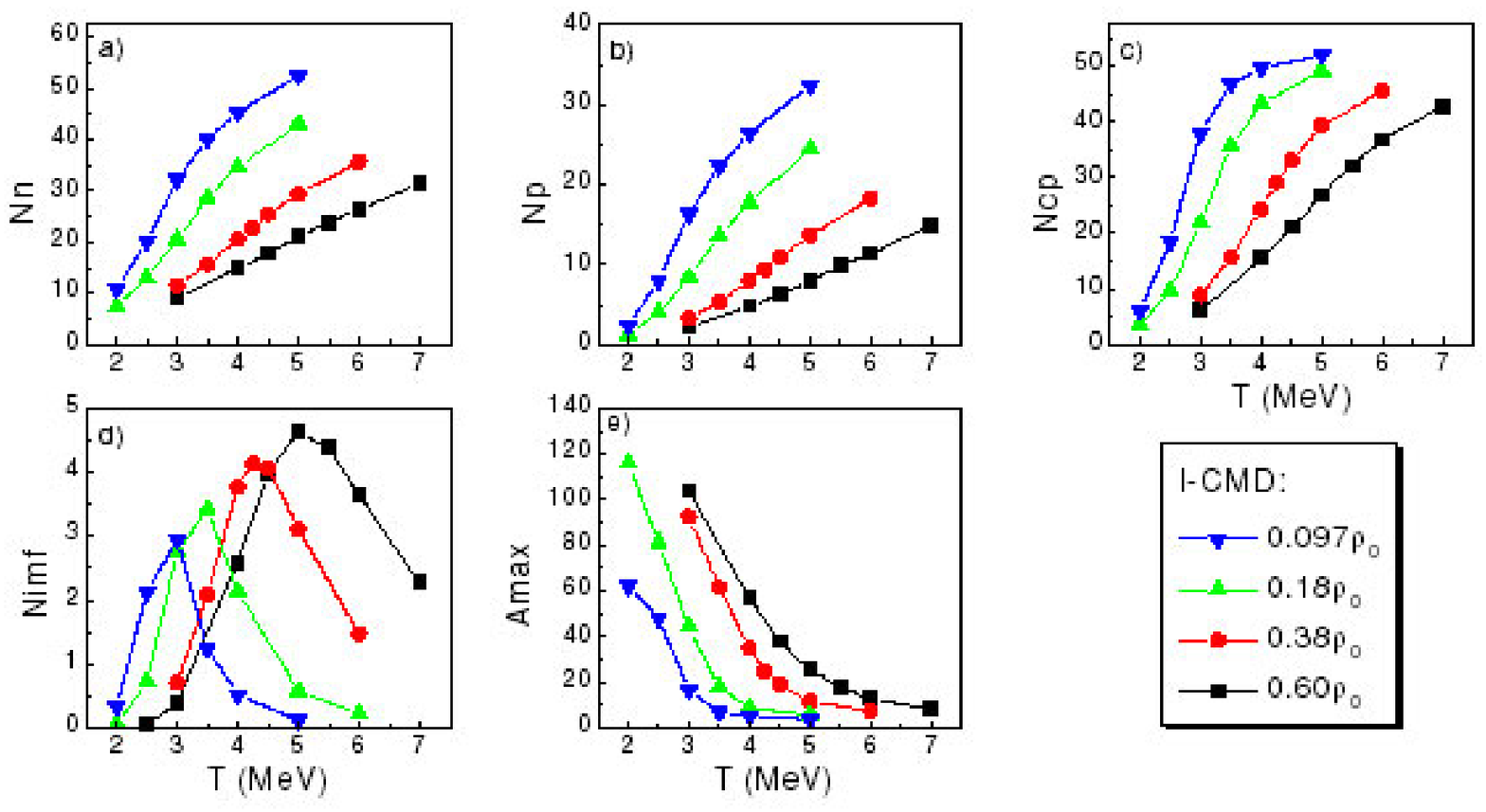}
\caption{\footnotesize Same as in figure 3 but in the case of I-CMD with Coulomb. }
\label{fig12}
\end{figure}  

\begin{figure}
\includegraphics[scale=0.43]{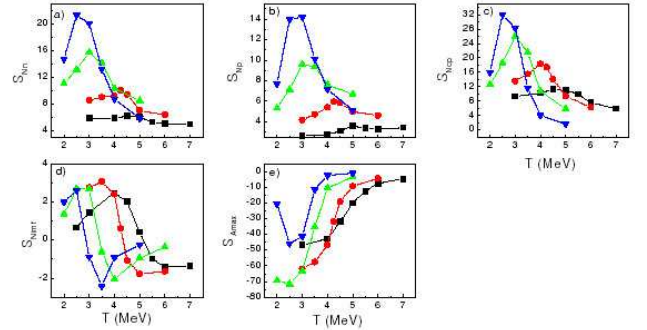}
\caption{\footnotesize  Same as in figure 4 but in the case of I-CMD with Coulomb. 
without the Coulomb. Thanks to this agreement, we treated the nuclear disassembly more 
realistically with switching on of the Coulomb interaction in the frame of I-CMD afterwards. }
\label{fig13}
\end{figure}  

\subsection{I-CMD in different 'freeze-out' densities }

Similar to the I-LGM cases, we also study the cluster emission in I-CMD in a wide range of 
`freeze-out' density, namely, 0.097$\rho_0$ ,0.18$\rho_0$ ,0.38$\rho_0$ and 
0.60$\rho_0$ to check phase transition behaviour. 

In order to compare it with the I-LGM case, similar figures are plotted in figures 12--15 
and compared to figures 3--6. Figure 12 shows that the mean multiplicities of emitted neutrons, 
protons, CP, IMF and $A_{max}$ evolve with temperature at different `freeze-out' densities in the 
I-CMD calculation with the Coulomb. Their emission rates or slopes and RMS width with 
temperature are depicted in figures 13 and 15. The effective power-law parameter of fragment 
distribution $\tau$ and multiplicity information entropy $H$ is plotted in figure 14. All these figures 
show the same behaviours as the I-LGM case, i.e. cluster emission rate and their fluctuation 
can provide us with the temperature of liquid--gas phase transition, namely around 3, 3.5, 4.5, 
5 MeV at $\rho_f$ = 0.097$\rho_0$ ,0.18$\rho_0$ ,0.38$\rho_0$ and 0.60$\rho_0$ respectively. 

Similarly, the Campi's scattering plots indicate the onset of liquid--gas phase transition 
around the respective transition temperatures. Figure 16 shows an example for I-CMD with 
the Coulomb. Clearly, phase coexistence takes place around 4.25 MeV. Overall, these phase 
transition temperatures are consistent with those extracted from the choppy position of the 
slopes of figure 13. Again, the largest fluctuation simultaneously appears at the point of phase 
transition. 

Overall, the phase transition temperature seems to rely on some ingredients, such as the 
`freeze-out' density (or pressure), the model and its interaction potential, but the rule that 
emission rate and fluctuation of cluster multiplicity can be taken as a probe of phase transition 
has not changed. 

\begin{figure}
\includegraphics[scale=0.43]{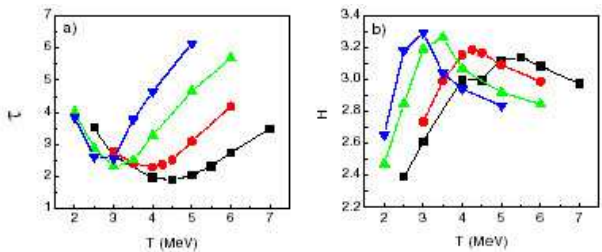}
\caption{\footnotesize  Same as in figure 5 but in the case of I-CMD with Coulomb. }
\label{fig14}
\end{figure}   

\begin{figure}
\includegraphics[scale=0.43]{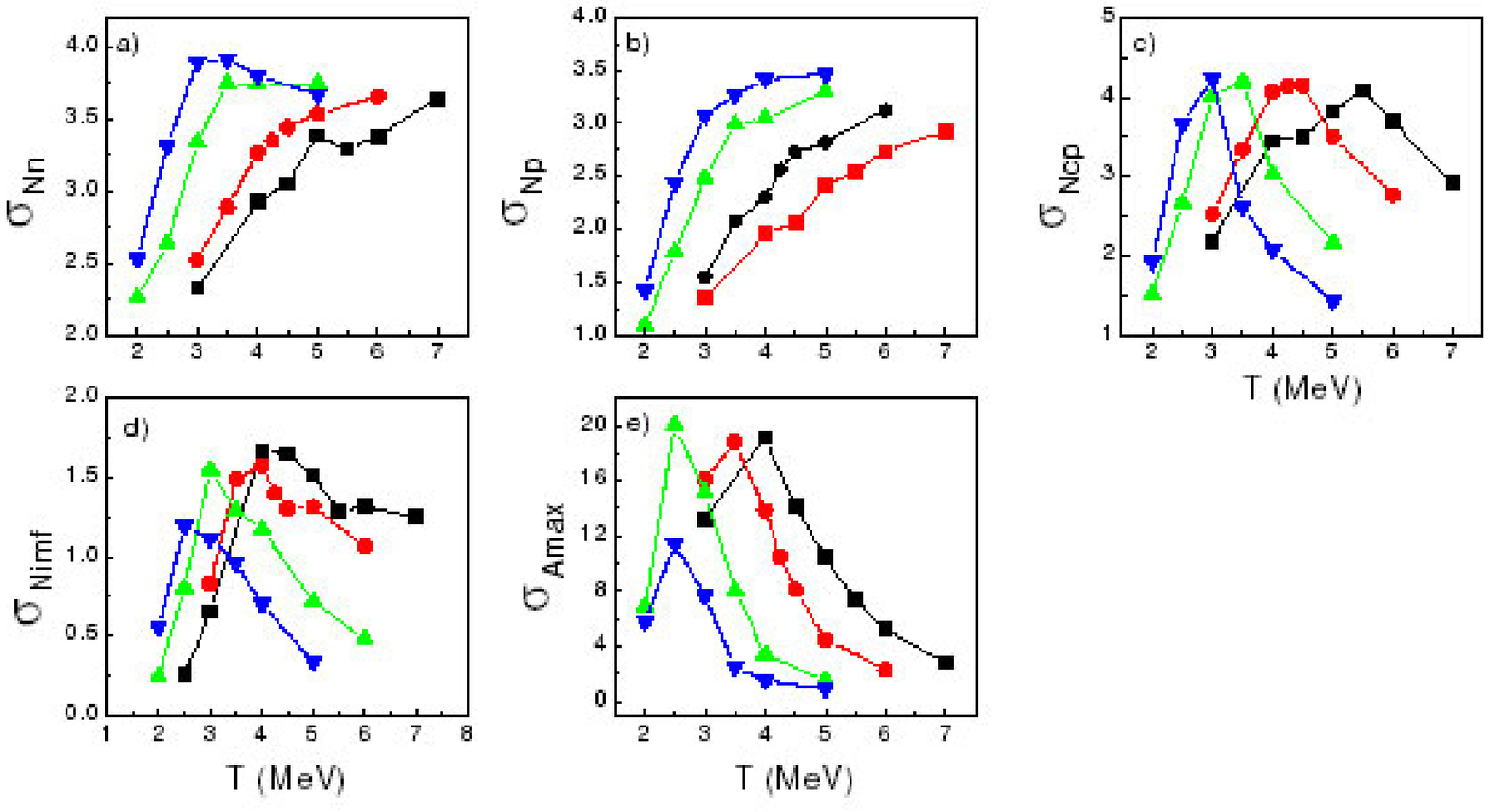}
\caption{\footnotesize Same as in figure 6 but in the case of I-CMD with Coulomb.  }
\label{fig15}
\end{figure}   

\begin{figure}
\includegraphics[scale=0.43]{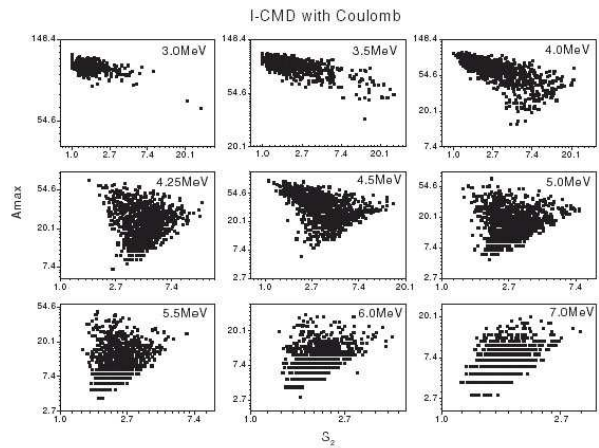}
\caption{\footnotesize Same as in figure 7, but for the I-CMD with Coulomb case at 0.38$\rho_0$ . }
\label{fig16}
\end{figure}   

\section{Conclusions }

In conclusion, the features of the emissions of LP, CP, IMF and MAX are investigated in 
a wide range of `freeze-out' density for a medium size nucleus $^{129}Xe$ in the frameworks of 
I-LGM and I-CMD model. $N_n$, $N_p$, $N_{cp}$ and $A_{max}$ show monotonous increase or decrease while 
$N_{imf}$ shows rise and fall with temperature. Slopes of these observables versus temperature go 
through extrema at the same temperature where the largest fluctuation of cluster multiplicity 
distributions is observed. This temperature is consistent with the phase transition temperature 
extracted from the extreme values of effective power law parameter $\tau$ and information entropy 
$H$ as well as Campi scatter plots. It gives an indication that the cluster emission rate can 
be taken as a probe of the phase transition of nuclei and furthermore, the largest fluctuation 
is simultaneously accompanied when the onset of phase transition occurs. In addition, the 
systematic comparison of I-LGM and I-CMD shows that LGM is a good tool to study nuclear 
disassembly when the system is not large where the Coulomb interaction can be ignored. But 
for large nuclear systems, I-CMD should be used to treat the nuclear dissociation and phase 
transition due to the importance of the Coulmb interaction. In light of this study, we think that 
the experimental study of cluster emission is rather meaningful, especially in measuring the 
excitation function of the multiplicities, their slopes and variances, from which some signals 
of phase transition could be found.

\acknowledgments 

I thank Prof S Das Gupta and Dr J Pan for kindly providing the original codes. I also 
appreciate Prof B Tamain and Prof J B Natowitz for their help. This study was partly 
supported by the National Natural Science Foundation of China for the Distinguished Young 
Scholar under Grant no 19725521, the National Natural Science Foundation of China under 
Grant no 19705012, the Science and Technology Development Foundation of Shanghai under 
grant no 97QA14038 and the Major State Basic Research Development Program of China 
under contract no G200077400. 

{}
\end{document}